\documentclass[12pt,reqno,a4paper]{amsart}

\usepackage{lipsum,subeqnarray}
\usepackage{amsfonts}
\usepackage{graphicx}
\usepackage{amsmath}
\usepackage{mathtools}
\usepackage{amssymb}
\usepackage{enumerate}
\usepackage{slashed}
\usepackage{graphicx}
\usepackage{newlfont}
\usepackage{amsrefs}
\usepackage{comment}
\usepackage[normalem]{ulem}
\usepackage{mathtools}
\usepackage{accents}
\usepackage{leftidx}
\usepackage{bbm}
\usepackage{amsthm}
\usepackage{appendix}

\numberwithin{equation}{section}

\theoremstyle{plain}
\newtheorem{theorem}{Theorem}[section]

\newtheorem*{theorem*}{Conjecture}

\usepackage{xcolor}

\theoremstyle{definition}

\theoremstyle{remark}
\newtheorem{remark}{Remark}
\usepackage{stackengine}
\usepackage{subeqnarray} 
\usepackage{esint}

\begin{document}
\title[Conformal deformations to strict DEC using a spacetime Poisson equation]{Conformal deformations of initial data sets to the strict dominant energy condition using a spacetime Poisson equation}
 
\author{Jaroslaw S. Jaracz}

\begin{abstract}
We give an alternate proof of one of the results given in \cite{Lee2022} showing that initial data sets with boundary for the Einstein equations $(M, g, k)$ satisfying the dominant energy condition can be conformally deformed to the strict dominant energy condition, while preserving the character of the boundary (minimal, future trapped, or past trapped) while changing the area of the boundary and ADM energy of the initial data set by an arbitrarily small amount. The proof relies on solving an equation that looks like the equation for spacetime harmonic functions studied in \cite{HirschKazarasKhuri}, but with a Neumann boundary condition and non-zero right hand side, which we refer to as a spacetime Poisson equation. One advantage of this method of proof is that the conformal deformation is explicitly constructed as a solution to a PDE, as opposed to only knowing the solution exists via an application of the implicit function theorem as in \cite{Lee2022}. We restrict ourselves to the physically relevant case of a $3$-manifold $M$, though the proof can be generalized to higher dimensions.  

\end{abstract}
\maketitle

\section{Introduction}

When studying initial data sets $(M, g, k)$ for the Einstein equations, one is usually interested in initial data sets satisfying certain energy conditions, such as the dominant energy condition
\begin{align*}
    \mu \geq |J|_g
\end{align*}
(see Section \ref{SecConstraintDEC} for definitions of the given quantities) which can be interpreted as saying that the speed of energy flow never exceeds that of the speed of light. 
However, for various technical reasons, it is easier to deal with the strict dominant energy condition
\begin{align*}
    \mu > |J|_g.
\end{align*}
The strict condition gives enough wiggle room to make certain technical arguments a lot easier. Therefore, when proving theorems for which the dominant energy condition is valid, one would like to to say it is possible to approximate the initial data sets by ones satisfying the strict condition in some appropriate sense.

The first example of this sort of argument was given in \cite{SchoenYau2}, and the deformation to the strict energy condition was claimed in Lemma 1 of that paper, using the implicit function theorem. However, as was pointed out in \cite{EichmairHuangLeeSchoen2016}, there was actually a flaw in this argument as the implicit function theorem was not applicable in that case. 

It was only in \cite{EichmairHuangLeeSchoen2016} that a correct argument was given (for manifolds without boundary), by introducing something called the modified Lie derivative, which allowed for the correct application of the implicit function theorem on Banach spaces. This was then generalized to the case of manifolds with boundary in \cite{Lee2022}. Now there are two things regarding these proofs which are not ideal. One is that the introduction of the modified Lie derivative and subsequent calculations can be somewhat tricky to follow along. The second is that since the proofs rely on the implicit function theorem, they only say that a deformation exists, but not how to obtain it. 

While studying conformal deformations for applications to the Penrose inequality (see Section \ref{Conclusion}), we stumbled across an alternative proof of such deformations, which does not rely on the implicit function theorem. On a conceptual level, the deformation is given as a solution to a certain quasi-linear elliptic PDE with Neumann boundary conditions. This also potentially has some useful practical applications. For example, if one wants to numerically study some initial data set, and wants to make sure the initial data set satisfies the strict dominant energy condition, one can explicitly calculate the solution to the aforementioned PDE. We thought this was sufficiently interesting from both the theoretical and practical point of view to be worth writing up.   

We do mention one advantage of the proofs in \cites{EichmairHuangLeeSchoen2016, Lee2022} is that they yield so-called harmonic asymptotics (which are in themselves useful), and it is unclear if our proof can be modified to yield these asymptotics as well. They also give results for initial data sets with lower regularity, though our proof should generalize to this case in a straightforward way.

We focus on the physically relevant case where $M$ is $3$-dimensional, though the ideas can be generalized to arbitrary dimension. To simplify presentation, we will assume that $M$ has asymptotially flat coordinates (see Section \ref{SubSecAFCoordinates}) in which
\begin{align} \label{ChoiceOfAFCoordinates}\begin{split}
    M=\mathbb{R}^3 \setminus U \\
    0 \notin M \end{split}
\end{align}
where $U$ is some domain in $\mathbb{R}^3$ with possibly several components, with $\partial U =\partial M$ smooth and compact with finitely many components. By choosing the coordinates so that $0\notin M$ we have $r(x)\geq c_0>0$ for all $x\in M$ and so we can take some constant $C$ such that \eqref{GeneralFallOffConditions} holds for all $x \in M$. 

This is done for convenience and to make the idea of the proof more transparent. The case of primary interest is that of one asymptotically flat end where $\partial M$ is nonempty and an outermost apparent horizon (defined below). In that case using a change of coordinates we can always choose coordinates such that \eqref{ChoiceOfAFCoordinates} holds. 

The theorem still applies to $M$ not of the form \eqref{ChoiceOfAFCoordinates} but in that case the function $r(x)$ appearing in \eqref{MainNeumannProblem} has to be replaced by some appropriately defined weight function $\sigma(x)$. This $\sigma(x)$ could be taken to be some smooth function defined to satisfy $\sigma\equiv 1$ on some compact $K$ and $\sigma(x)=r(x)$ sufficiently far in the asymptotically flat end. All the subsequent proofs can be generalized to this case in a straightforward, though slightly messy manner.

The main theorem we prove in this paper is:

\begin{theorem} \label{MainTheorem}
    Let $(M, g, k)$ be a smooth asymptotically flat initial data set for the Einstein equations with $M$ of the form \eqref{ChoiceOfAFCoordinates}, satisfying the dominant energy condition, with one asymptotically flat end and compact boundary $\partial M$ of area $A$, satisfying the fall-off conditions \eqref{GeneralFallOffConditions} with constant $C$, and with ADM energy $E$. Then there exists a $\psi \in C^{2, \alpha}_{-1}({M})$ solving the Neumann problem
    \begin{align} \label{MainNeumannProblem}
        \begin{split}
            \Delta_g \psi + \frac{C}{r^2} \vert \nabla \psi \vert_g &= -\frac{1}{r^4} \\
            \left.\frac{\partial \psi}{\partial \nu}\right\vert_{\partial M} &= 0.
        \end{split}
    \end{align}
Moreover, if we define $u_\epsilon=1+\epsilon \psi$ for all sufficiently small $\epsilon > 0$ then the initial data set $(M, g_\epsilon, k_\epsilon)=(M, u_\epsilon^4 g, u_\epsilon^2 k)$ is asymptotically flat, satisfies the strict dominant energy condition, and there exists some constant $\mathcal{C}>0$ such that 
\begin{align} \label{EnergyAreaEstimate}
    \vert E - E_\epsilon| \leq \mathcal{C}\epsilon, \quad \vert A-A_\epsilon \vert \leq \mathcal{C}\epsilon
\end{align}
where $E_\epsilon$ and $A_\epsilon$ are the ADM energy and boundary area of $g_\epsilon$. Moreover, the sign of the null expansions at the boundary is the same for $g$ and $g_\epsilon$, so in particular apparent horizon boundaries remain apparent horizons. 
\end{theorem}

We mention that the right hand side of \eqref{MainNeumannProblem} can be replaced by any negative function which vanishes sufficiently fast at infinity, giving further (potentially useful) control over the deformation. The proof closely follows \cite{HirschKazarasKhuri} where a similar equation, but with Dirichlet boundary condition, is studied. The proof consists of the application of standard tools, such as the Leray-Schauder fixed point theorem, maximum principles, and some simple elliptic estimates.

Finally, we mention that a linear elliptic equation with Neumann boundary condition was used to study the scalar curvature of initial data sets $(M, g)$ in \cite{Lee2023}.

\section{Background Material}

\subsection{Asymptotic Flatness and ADM Energy}\label{SubSecAFCoordinates}

 We consider an asymptotically flat initial data set $(M, g, k)$ where $M$ is a $3$-manifold, $g$ a Riemannian metric, and $k$ is a symmetric $2$-tensor, the extrinsic curvature. We assume the manifold is smooth and has a single asymptotically flat end and some smooth compact boundary $\partial M$ with possibly multiple components. These components are not necessarily apparent horizons, though the apparent horizon case is the one of greatest interest. We assume that $g$ and $k$ are of class $C^{\infty}({M})$. 
 
An initial data set having a single asymptotically flat end means that there is some compact set $K$ such that $M\setminus K$ is diffeomorphic to the complement of a ball in $\mathbb{R}^3$. This diffeomorphic region is referred to as the asymptotically flat end. It is assumed that on this asymptotically flat end $g$ and $k$ satisfy certain fall-off conditions. In our case, we take these to be
\begin{align} \label{GeneralFallOffConditions} 
    | D^{\lambda} ( g_{ij} - \delta_{ij}) | \leq Cr^{-1-|\lambda|}, \quad |R|\leq Cr^{-3}, \quad |D^\lambda k|\leq Cr^{-2-|\lambda|}
\end{align}
for $|\lambda|\leq 2$. Here, $C$ is some constant, $R$ is the scalar curvature, $\delta$ is the Euclidean metric, $r=\sqrt{x^2+y^2+z^2}$ the standard Euclidean radius, $D^\lambda$ is a derivative operator with respect to the Euclidean coordinates, and $\lambda$ is a multi-index. We have 
\begin{equation*}
    |k|^2=k_{ij}k^{ij}, \quad Tr_g k = g^{ij}k_{ij}
\end{equation*}
as usual, where the indices are raised with $g$.

For an asymptotically flat end, the ADM energy is defined by 
\begin{align} \label{ADMEnergy} 
E_{ADM}=\lim_{r\rightarrow \infty}\frac{1}{16\pi} \sum_{i, j} \int_{S_r} (g_{ij,i}-g_{ii,j})\nu^j dS_r 
\end{align}
where $S_r$ are coordinate spheres of radius $r$ and $\nu^j$ is the outward unit normal \cite{ADM}.

\subsection{Einstein Constraint Equations and Energy Conditions}\label{SecConstraintDEC}

The Einstein constraint equations for an initial data set $(M, g, k)$ are
\begin{align} \label{ConstraintEquations}
\begin{split}
\mu&=\frac{1}{2}\left(R+(Tr_gk)^2-|k|_g^2\right)\\
J_i&=\nabla^j (k_{ij}-(Tr_g k)g_{ij})
\end{split}
\end{align}
where $R$ is the scalar curvature, $\nabla^j$ denotes covariant differentiation, $\mu$ is the mass-energy density, and $J_i$ the components of the momentum density. Then the dominant energy condition (DEC) is that the inequality
\begin{equation}\label{DEC}
    \mu \geq |J|_g 
\end{equation}
is satisfied everywhere in the manifold.

\subsection{Null expansions and apparent horizons}

Given a (two sided) codimension-$1$ hypersurface in $M$ one can calculate the future $(+)$ and past $(-)$ null expansions at each point of the surface, which are defined by
\begin{align}
    \theta_\pm = H \pm Tr_S k.
\end{align}
Here $H$ is the mean curvature computed with respect to the unit normal $\nu$ pointing towards spatial infinity in a chosen AF end and $Tr_S k$ denotes the trace of $k$ restricted to the surface $S$ computed with respect to the induced metric on the surface. 

The futre and past null expansions measure the convergence and divergence of past and future directed null geodesics. Surfaces with $\theta_\pm <0$ are referred to as future and past trapped. Surfaces with 
$\theta_\pm$ are referred to as future and past apparent horizons and are a popular way of modelling the surface of a black hole in an initial data set.

\subsection{Conformal Change}

A conformal change of an initial data set $(M, g, k)$ to a new initial data set $(M, \bar{g}, \bar{k})$ is defined by
\begin{align*}
    (M, g, k) \rightarrow (M, \overline{g}, \overline{k})=(M, u^4 g, u^2 k).
\end{align*}
Here $u>0$ is some non-negative function on $M$. It is then useful to know how the relevant quantities transform under this conformal change. One has
\begin{align} \label{muJconformalchange}\begin{split}
    \overline{\mu}&= u^{-4} \left( \mu -4u^{-1} \Delta_g u   \right) \\
    \overline{J}_i &= u^{-2} \left(  J_i +4u^{-1}  k_{ij} \nabla^j u\right). \end{split}
\end{align}
See the formulas in Lemma 1 of \cite{SchoenYau2} where our $\overline{J}_i$ is related to the $K_i$ defined there by $\overline{J}_i=u^{-2}K_i$. Though the proof of the lemma incorrectly applies the implicit function theorem on Banach spaces, the formulas for conformal change are correct.

We want to derive some condition on $u$ such that the DEC is preserved. Notice that for a $1$-form $X_i$ we have
\begin{align}
    |X_i|_{\overline{g}}^2 = \overline{g}^{ij}X_i X_j = u^{-4} g^{ij}X_i X_j = u^{-4}|X_i|_g^2
\end{align}
so that
\begin{align}
    \vert X_i \vert_{\overline{g}} = u^{-2}\vert X_i \vert_g.
\end{align}
Combining this with \eqref{muJconformalchange} and the triangle inequality we obtain 
\begin{align}
    \vert \overline{J}_i\vert_{\overline{g}} \leq \vert u^{-2} J_i \vert_{\overline{g}} + \vert 4u^{-3}  k_{ij} \nabla^j u \vert_{\overline{g}} = u^{-4} \vert  J_i \vert_{{g}} + 4u^{-5}\vert k_{ij} \nabla^j u \vert_{{g}}
\end{align}
Therefore, if we have 
\begin{align} 
    \Delta_g u \leq  -|k_{ij}\nabla^j u|_g
\end{align}
then 
\begin{align} \label{estimatemuJ}
      \overline{\mu}&= u^{-4} \mu -4u^{-5} \Delta_g u \geq u^{-4} \mu + 4 u^{-5} |k_{ij} \nabla^j u|_g \geq u^{-4} \vert  J_i \vert_{{g}} + 4u^{-5}\vert k_{ij} \nabla^j u \vert_{{g}} \geq  \vert \overline{J}_i\vert_{\overline{g}}
\end{align}
and the dominant energy condition is preserved. Of course, then the most natural equation to try to solve is 
\begin{align} \label{ImportantEquation}
    \Delta_g u =  -|k_{ij}\nabla^j u|_g.
\end{align}
However, since using \eqref{GeneralFallOffConditions} we can estimate
\begin{align} \label{Estimatek}
    |k_{ij}\nabla^j u|_g \leq \frac{C}{r^2}|\nabla u|_g
\end{align}
then if we look to solve
\begin{align}\label{MainEquation1}
    \Delta_g u = -\frac{C}{r^2}|\nabla u|_g
\end{align}
then we get 
\begin{align*}
    \Delta_g u = -\frac{C}{r^2}|\nabla u|_g \leq -|k_{ij} \nabla^j u|_g
\end{align*}
preserving the dominant energy condition. Notice that \eqref{MainEquation1} is of the same form as the equation for spacetime harmonic functions for which existence results were established in \cite{HirschKazarasKhuri}, which we will exploit. The equation studied in \cite{HirschKazarasKhuri} was of the form
\begin{equation} \label{SpacetimeLaplaceEquation}
    \Delta_g u + G(x) |\nabla u|_g = 0
\end{equation}
and since the solutions were referred to as spacetime harmonic functions, it makes sense to refer to \eqref{SpacetimeLaplaceEquation} as a spacetime Laplace equation. Therefore it also makes sense to refer to an equation of the form 
\begin{align}
\label{SpacetimePoissonEquation}
    \Delta_g u + G(x) |\nabla u|_g = F(x)
\end{align}
as a spacetime Poisson equation, hence our terminology. 

Let $S$ be a two sided codimension-$1$ surface in $M$ with mean curvature $H$ computed with respect to $g$ with the chosen unit normal $\nu$. Then the mean curvature computed with respect to $\bar{g}$ with respect to the unit normal $\overline{\nu}=u^{-2}\nu$ is given by the standard formula
\begin{align}
    \overline{H}=u^{-2}\left( H + 4 u^{-1} \langle \nabla u, \nu \rangle_g  \right),
\end{align}
see formula (7) in \cite{EichmairHuangLeeSchoen2016}.

If we denote by $h$ the induced metric on $S$ then $\overline{h}=u^4 h$. If we let $K=k_S$ denote $k$ restricted to $S$, and similarly $\overline{K}=\overline{k}_S$ then we get 
\begin{align}
    \overline{K}=u^2 K
\end{align}
so that
\begin{align}
    Tr_S \overline{k} = \overline{h}^{ij} \overline{K}_{ij}=u^{-4}h^{ij} u^2 K_{ij} = u^{-2} h^{ij} K_{ij} = u^{-2} Tr_S k.
\end{align}
Therefore we have
\begin{align}\label{NullExpansionUnderConformalChange}
    \overline{\theta}_\pm = u^{-2} \theta_\pm +4u^{-3}\langle \nabla u, \nu \rangle_g
\end{align}
by combining the above formulas.

\section{Proof of Theorem \ref{MainTheorem} }

We first prove all the claims of the theorem assuming a solution to \eqref{MainNeumannProblem} exists. We then prove the existence of $\psi$. For the definitions of the weighted H\"{o}lder spaces $C^{k, \alpha}_{-\beta}$ we refer the reader to \cites{EichmairHuangLeeSchoen2016, Chaljub1979, Weinstein_2005}. Suppose a solution $\psi\in C^{2, \alpha}_{-1}({M})$ to the problem \eqref{MainNeumannProblem} exists. Then in particular $\psi$ is bounded.  Define
\begin{align}
    u_\epsilon = 1+ \epsilon \psi.
\end{align} so for all sufficiently small $\epsilon>0$ we have that $1+\epsilon \psi >0$ so $g_\epsilon=u_\epsilon^4 g$ is a Riemannian metric. Due to the fact that $\psi \in C^{2, \alpha}_{-1}({M})$ the function $u_\epsilon$ tends towards $1$ as $|x|\rightarrow \infty$ and the derivatives $\partial_i u_\epsilon$ and $\partial_{ij} u_\epsilon$ vanish as $O(r^{-2})$ and $O(r^{-3})$ respectively. Therefore, also defining $k_\epsilon=u_{\epsilon}^2 k$ yields that $(M, g_\epsilon, k_\epsilon)$ 
is an asymptotically flat initial data set.

Next, notice that 
\begin{align} \begin{split}
    \Delta_g u_\epsilon + \frac{C}{r^2} \vert \nabla u_\epsilon \vert_g &= \Delta_g (\epsilon\psi) + \frac{C}{r^2} \vert \nabla (\epsilon\psi) \vert_g = \epsilon \Delta_g \psi + \epsilon\frac{C}{r^2} \vert \nabla \psi \vert_g \\&= \epsilon \left( \Delta_g \psi + \frac{C}{r^2} \vert \nabla \psi \vert_g \right)   = -\frac{\epsilon}{r^4} \end{split}
\end{align}
since $\psi$ satisfies \eqref{MainNeumannProblem} and we were able to move $\epsilon$ through the norm since $\epsilon>0$. Then
\begin{align} \begin{split}
      \mu_\epsilon&= u_\epsilon^{-4} \mu -4u_\epsilon^{-5} \Delta_g u_\epsilon = u_\epsilon^{-4} \mu +4u_\epsilon^{-5} \left( \frac{C}{r^2} |\nabla u_\epsilon|_g + \frac{\epsilon}{r^4}  \right) \\
      &>  u_\epsilon^{-4} \mu_\epsilon + 4 u_\epsilon^{-5} |k_{ij} \nabla^j u_\epsilon|_g \geq u_\epsilon^{-4} \vert  J_i \vert_{{g}} + 4u_\epsilon^{-5}\vert k_{ij} \nabla^j u_\epsilon \vert_{{g}}\\ &\geq  \vert (J_\epsilon)_i\vert_{g_\epsilon} \end{split}
\end{align}
where we used \eqref{muJconformalchange},\eqref{estimatemuJ}, and \eqref{Estimatek}. So indeed, the conformal initial data set satisfies the strict dominant energy condition. Moreover, we can make the estimate
\begin{align}
    \mu_\epsilon - |J_\epsilon|_{g_\epsilon} \geq \frac{4\epsilon u_\epsilon^{-5}}{r^4} >0
\end{align}
and since for sufficiently small $\epsilon$ we have that $u_\epsilon$ is close to $1$ we have
\begin{align}
    \mu_\epsilon - |J_\epsilon|_{g_\epsilon} \geq \frac{\epsilon}{r^4} >0
\end{align}
for all sufficiently small $\epsilon>0$.

The estimates \eqref{EnergyAreaEstimate} follow easily from the fact that $\psi \in C^{2, \alpha}_{-1}({M})$ and the formula \eqref{ADMEnergy} for the ADM energy. Finally, since $\psi$ satisfies the Neumann condition at the boundary, \eqref{NullExpansionUnderConformalChange} shows that
\begin{align}
    (\theta_\epsilon)_\pm = u_\epsilon^{-2} \theta_\pm
\end{align}
on $\partial M$ so the sign of the null expansions is the same for any $p \in \partial M$.

\subsection{Solution of \eqref{MainNeumannProblem} on compact subdomains} The existence of a solution to \eqref{MainNeumannProblem} is established using similar methods to Section 4.1 in \cite{HirschKazarasKhuri} for the case of spacetime harmonic functions. Let $S_{\mathcal{R}}$ denote a coordinate sphere of radius $\mathcal{R}$ in $M$. Let $M_{\mathcal{R}}$ be the compact set defined by
\begin{align}
    M_{\mathcal{R}}=\lbrace x\in M: r(x)\leq \mathcal{R} \rbrace
\end{align}which has boundary $\partial M_{\mathcal{R}}=\partial M \cup S_{\mathcal{R}}$. Consider the mixed boundary value problem
\begin{align} \label{MixedBoundaryValueProblem}
        \begin{split}
            \Delta_g \psi_{\mathcal{R}} + \frac{C}{r^2} \vert \nabla \psi_{\mathcal{R}} \vert_g &= -\frac{1}{r^4} \quad \text{on} \quad M_{\mathcal{R}} \\
            \left.\frac{\partial \psi_{\mathcal{R}}}{\partial \nu}\right\vert_{\partial M} = 0, \quad &\psi_{\mathcal{R}}\vert_{S_{\mathcal{R}}}=0
        \end{split}
    \end{align}
Recall we chose the coordinates \eqref{ChoiceOfAFCoordinates} ensuring $r(x)\geq c_0>0$ for some constant $c_0$ so that all the coefficients are smooth. We will solve this subproblem using the Leray-Schauder fixed point theorem, see Theorem 11.3 in \cite{GilbargTrudinger}:

\begin{theorem}[Leray-Schauder Fixed Point Theorem]
    Let $\mathcal{B}$ be a Banach space and $\mathcal{F} : \mathcal{B} \times [0, 1] \rightarrow \mathcal{B}$ a compact mapping with $\mathcal{F}(b, 0) = 0$
for all $b\in \mathcal{B}$. If there is a constant $c$, such that any solution $(b, \sigma) \in \mathcal{B} \times[0, 1]$ of $b = \mathcal{F}(b, \sigma)$ satisfies
the a priori inequality $\Vert b \Vert_{\mathcal{B}} \leq c$, then there is a fixed point at $\sigma=1$. That is, there exists $b_1 \in \mathcal{B}$ with
$b_1=\mathcal{F}(b_1, 1)$.
\end{theorem}
 
Define the space
\begin{align}
    \widetilde{C}^{k, \alpha} ({M}_{\mathcal{R}}) =  \lbrace f\in {C}^{k, \alpha} ({M}_{\mathcal{R}}): \partial f/ \partial \nu \vert_{\partial M}=0, \quad f\vert_{S_{\mathcal{R}}}=0 \rbrace
\end{align}
consisting of functions satisfying the Neumann boundary condition on $\partial M$ and the zero Dirichlet boundary condition on $S_{\mathcal{R}}$. Notice that $\Delta^{-1}_g: C^{0, \alpha} \rightarrow  \widetilde{C}^{2, \alpha}$ is an isomorphism by applying the Fredholm alternative in the usual way. Next, define the operator
\begin{align}
    F(w)= -\frac{C}{r^2} \vert \nabla w  \vert_g  -\frac{1}{r^4}
\end{align}
and notice if $w\in \widetilde{C}^{1, \alpha}(M_{\mathcal{R}})$ then $F(w)\in C^{0, \alpha}(M_{\mathcal{R}})$. Letting $\mathcal{B}= \widetilde{C}^{1, \alpha}(M_{\mathcal{R}})$ we define $\mathcal{F}(w, \sigma):\mathcal{B} \times [0, 1]\rightarrow \mathcal{B}$ to be the composition of maps
\begin{align}
    \widetilde{C}^{1, \alpha}(M_{\mathcal{R}})\xrightarrow{F} C^{0, \alpha}(M_{\mathcal{R}}) \xrightarrow{\sigma \Delta_g^{-1}} \widetilde{C}^{2, \alpha}(M_{\mathcal{R}}) \xrightarrow{i} \widetilde{C}^{1, \alpha}(M_{\mathcal{R}})
\end{align}
which is a compact map since the first two maps are bounded and the last inclusion is compact. Notice, $v=\mathcal{F}(w, \sigma)$ if $v$ solves the problem
\begin{align} 
        \begin{split}
            &\Delta_g v = \sigma \left(- \frac{C}{r^2} \vert \nabla w \vert_g -\frac{1}{r^4}\right) \\
            &\left.\frac{\partial v}{\partial \nu}\right\vert_{\partial M} = 0, \quad v\vert_{S_{\mathcal{R}}}=0.
        \end{split}
    \end{align}
We denote the fixed points of the mapping by $w_\sigma=\mathcal{F}(w_\sigma, \sigma)$. Notice,  $\mathcal{F}(w, 0)=0$. Next, we want to obtain an estimate of the form
\begin{align}
    \Vert w_\sigma \Vert_{\widetilde{C}^{1, \alpha}(M_{\mathcal{R}})} \leq c
\end{align}
for all $0\leq \sigma \leq 1$. The key is to notice that we can always rewrite
\begin{align}
    \vert \nabla w \vert_g = \frac{g_{ij}\nabla^j w}{|\nabla w|_g} \nabla^i w
\end{align}
where we can think of ${g_{ij}\nabla^j w}/{|\nabla w|_g}$ as a coefficient in $L^\infty$. To be slightly more precise, this is undefined when $|\nabla w|_g=0$ so we should define a coefficient depending on $\nabla w$ by
\begin{align}
    b_i(\nabla w, x) \coloneqq \begin{dcases}
        \frac{g_{ij}\nabla^j w}{|\nabla w|_g}  \quad &: \quad |\nabla w|_g \neq 0 \\
        0 \quad &: \quad |\nabla w|_g=0
    \end{dcases}.
\end{align}
In this case indeed 
\begin{align}
    b_i(\nabla w, x)\nabla^i w =\begin{dcases}
        {|\nabla w|_g}  \quad &: \quad |\nabla w|_g \neq 0 \\
        0 \quad &: \quad |\nabla w|_g=0 
    \end{dcases} \quad = |\nabla w|_g
\end{align}
and $|b_i(\nabla w, x)|\leq c_1$ are uniformly bounded independent of $\nabla w$. To improve readability, we write $ b_i(\nabla w, x)=b_i(\nabla w)$

Therefore, the fixed points of $\mathcal{F}$ satisfy the problem
\begin{align} 
        \begin{split}
            &\Delta_g w_\sigma = \sigma \left(- \frac{C}{r^2}  b_i(\nabla w_\sigma) \nabla^i w_\sigma -\frac{1}{r^4}\right) \\
            &\left.\frac{\partial w_\sigma}{\partial \nu}\right\vert_{\partial M} = 0, \quad w_\sigma\vert_{S_{\mathcal{R}}}=0.
        \end{split}
    \end{align}
Therefore, if a solution exists, it can be viewed as a solution to the linear problem
\begin{align} 
        \begin{split}
            &\Delta_g v = \sigma \left(- \frac{C}{r^2} b_i(\nabla w_\sigma) \nabla^i v -\frac{1}{r^4}\right) \\
            &\left.\frac{\partial v}{\partial \nu}\right\vert_{\partial M} = 0, \quad v\vert_{S_{\mathcal{R}}}=0.
        \end{split}
    \end{align}
and this linear problem has uniformly bounded coefficients for all $0\leq \sigma \leq 1$, independent of $w_\sigma$. Therefore, we can apply the usual $L^p$ estimates for elliptic equations which give
\begin{align}
    \Vert w_\sigma \Vert_{W^{2, p}(M_{\mathcal{R}})} \leq C_1 \left( \Vert r^{-4} \Vert_{{L^{p}(M_{\mathcal{R}})}}  + \Vert w_\sigma \Vert_{L^{p}(M_{\mathcal{R}})}  \right)
\end{align}
for some elliptic constant $C_1$. Since the coefficient of the zero order term in the linear PDE is zero, the maximum principle can be applied which gives a $C^0$ and hence a $L^p$ bound for $w_\sigma$. Therefore we obtain an estimate
\begin{align}
    \Vert w_\sigma \Vert_{W^{2, p}(M_{\mathcal{R}})} \leq C_2
\end{align}
for some constant $C_2$ independent of $\sigma$. For sufficiently large $p$, the Sobolev embedding theorem gives ${W^{2, p}(M_{\mathcal{R}})} \hookrightarrow C^{1, \alpha}(M_{\mathcal{R}})$ so we have the estimate
\begin{align}
    \Vert w_\sigma \Vert_{C^{1, \alpha}(M_{\mathcal{R}})} \leq C_3
\end{align}
independent of $\sigma$. We can now apply the Leray-Schauder fixed point theorem for $\sigma=1$ to obtain a solution to \eqref{MixedBoundaryValueProblem} with at least $\widetilde{C}^{1, \alpha}(M_{\mathcal{R}})$ regularity which we denote by $w_1=\psi_{\mathcal{R}}$. But now we can look at it as as solution of 
\begin{align} \label{MixedBoundaryValueProblem2}
        \begin{split}
            \Delta_g \psi_{\mathcal{R}} &= f\quad \text{on} \quad M_{\mathcal{R}} \\
            \left.\frac{\partial \psi_{\mathcal{R}}}{\partial \nu}\right\vert_{\partial M} &= 0, \quad \psi_{\mathcal{R}}\vert_{S_{\mathcal{R}}}=0
        \end{split}
    \end{align}
where 
\begin{align} \label{fC0alpha}
    f=- \frac{C}{r^2} \vert \nabla \psi_{\mathcal{R}} \vert_g  -\frac{1}{r^4} \in C^{0, \alpha}(M_{\mathcal{R}}) 
\end{align}
so therefore $\psi_{\mathcal{R}} \in C^{2, \alpha}(M_{\mathcal{R}})$.

Next, we need to show that we can take the limit $\mathcal{R}\rightarrow \infty$. This can be done by constructing a pair of sub and super solutions to \eqref{MixedBoundaryValueProblem} which are independent of $\mathcal{R}$. This is done in the same way as in \cite{HirschKazarasKhuri}, except our case is even simpler due to the form of our equation.

\subsection{A uniform $C^0$ bound for $\psi_{\mathcal{R}}$}
Let us remind ourselves what the definitions of sub and supersolutions are. Given the equation
\begin{align}
    Lu&=f(x) \quad \text{on} \quad \Omega \\
    Bu&=h(x) \quad \text{on} \quad \partial \Omega 
\end{align}
where $L$ is a linear uniformly elliptic differential operator and $Bu=A(x) \partial u/\partial \nu + B(x)u$ with either $A=0, B>0$ or $A>0,  B\geq 0$ is a linear boundary operator, we say $\overline u$ is a supersolution if it satisfies 
\begin{align}
    L\overline{u}&\leq f(x) \quad \text{on} \quad \Omega \\
    B\overline{u}&\geq h(x) \quad \text{on} \quad \partial \Omega. 
\end{align}
Similarly we say $\underline u$ is a subsolution if it satisfies 
\begin{align}
    L\underline{u}&\geq f(x) \quad \text{on} \quad \Omega \\
    B\underline{u}&\leq h(x) \quad \text{on} \quad \partial \Omega. 
\end{align}
If $\overline{u}, \underline{u}\in H^1(\Omega)$ then we mean these inequalities to hold in the weak sense.

Since we have shown $\psi_{\mathcal{R}}$ exists for each $\mathcal{R}$ we can think of the family of linear problems with uniformly bounded coefficients
\begin{align}  \label{MixedProblemSubSuper}
        \begin{split}
            L_{\mathcal{R}}(v) \coloneqq &\Delta_g v + \frac{C}{r^2}  b_i(\nabla \psi_\mathcal{R}) \nabla^i v= -\frac{1}{r^4} \quad \text{on} \quad M_{\mathcal{R}} \\
            &\left.\frac{\partial v}{\partial \nu}\right\vert_{\partial M} = 0, \quad v\vert_{S_{\mathcal{R}}}=0.
        \end{split}
    \end{align}
and we would like to show there are sub and supersolutions to this problem which are independent of $\mathcal{R}$. 
In fact, by the definition of a subsolution we have $\underline{\psi}=0$ is a subsolution to \eqref{MixedProblemSubSuper} for any $\mathcal{R}$. It remains to construct a supersolution $\overline{\psi}$. 

Consider the function $\overline{w}(r)$ 
\begin{align} \label{DefinitionOverlineW}
    \overline{w}(r)=\lambda r^{-\beta}, \quad \overline{w}'(r)=-\lambda \beta r^{-1-\beta}, \quad \overline{w}''(r)=\lambda \beta (1+\beta)r^{-2-\beta}
\end{align}
for $\lambda>0$ and $\beta \in (0, 1)$. Using the fall-off conditions \eqref{GeneralFallOffConditions}, it can be shown that
\begin{align} \label{DeltagAsymptotics}
    \Delta_g \overline{w} = -\lambda \beta (1-\beta)r^{-2-\beta}\left( 1+O(r^{-1})    \right),
\end{align}
see equation (4.20) in \cite{HirschKazarasKhuri}. It can also be verified that this is correct up to the leading order term by calculating the Laplacian of $\overline{w}$ with respect to the Euclidean metric. 

Fix some $\beta\in (0, 1)$. Then for any $\lambda>1$ there exists some $r_0$ such that 
\begin{align} \label{SuperSolutionEstimate1} 
        \begin{split}
            &\Delta_g \overline{w} + \frac{C}{r^2} b_i(\nabla \psi_{\mathcal{R}}) \nabla^i \overline{w}\leq  -\frac{1}{r^4} \quad \text{for} \quad r\geq r_0
        \end{split}
    \end{align}
independent of $\mathcal{R}$ (because for $\beta \in (0,1)$ the coefficient in \eqref{DeltagAsymptotics} is negative and the leading term decays as $r^{-2-\beta}$). Similar to  \cite{HirschKazarasKhuri}, we define the function
\begin{align}
    \overline{\psi} = \begin{dcases}
        \widetilde{\psi} \coloneqq \widehat{\psi}_{r_0} + \lambda r_0^{-\beta} \quad &: \quad r\leq r_0 \\
       \overline{w}= \lambda r^{-\beta} \quad &: \quad r>r_0
    \end{dcases}
\end{align}
where $\widehat{\psi}_{r_0}$ is the solution of \eqref{MixedBoundaryValueProblem} on $M_{r_0}$ but with $-2r^{-4}$ instead of $-r^{-4}$ on the right hand side of the equation. 
Notice, $\lambda r_0^{-\beta}$ is just a constant. We see that $\overline{\psi}$ is $C^{2, \alpha}$ on $M_{r_0}$, smooth on $M\setminus M_{r_0}$ and by construction continuous on $S_{r_0}$ for any $\lambda>1$. It is therefore easy to show $\overline{\psi}\in H^1(M_{\mathcal{R}})$ for any $\mathcal{R}$.

Let $\partial_r f$ denote the outward normal derivative of $f$ on $S_{r_0}$. Due to \eqref{DefinitionOverlineW}, for a sufficiently large $r_0$ we can take a sufficiently large $\lambda>1$ such that 
\begin{align} \label{NormalDerivativeEstimate}
    \partial_r \widetilde{\psi} > \partial_r \overline{w} \quad \text{on} \quad S_{r_0}.
\end{align}
We can now show that $\overline{\psi}$ is a supersolution. Notice,
\begin{align}
    b_{i}(\nabla w)\nabla^i v \leq |\nabla v|_g
\end{align}
for any $w, v$. Therefore 
\begin{align} \label{SuperSolutionEstimate2} \begin{split}
    L_{\mathcal{R}}(\widetilde{\psi})&=\Delta_g \widetilde{\psi} + \frac{C}{r^2}b_i(\nabla \psi_{\mathcal{R}})\nabla^i \widetilde{\psi} \leq \Delta_g \widetilde{\psi} + \frac{C}{r^2}|\nabla \widetilde{\psi}|_g = \Delta_g \widehat{\psi}_{r_0} + \frac{C}{r^2}|\nabla {\widehat{\psi}_{r_0}}|_g = -\frac{2}{r^4} \\ &< -\frac{1}{r^4} \end{split}
\end{align}
on $M_{r_0}\cap M_{\mathcal{R}}$. Therfore $\overline{\psi}$ is a supersolution on $M_{r_0}\cap M_{\mathcal{R}}$ and $M_{\mathcal{R}}\setminus M_{r_0}$ separately. The problem is that the function is not differentiable across $S_{r_0}$. However, we can show that it is a weak supersolution which is enough.

For our problem, $\overline{\psi}$ will be a supersolution if for a non-negative test function $\phi \in C_c^{\infty}(M_{\mathcal{R}})$ we have
\begin{align}
    \int_{M_\mathcal{R}} \left(-\langle \nabla \phi, \nabla \overline{\psi} \rangle_g + \phi \frac{C}{r^2}b_i(\nabla \psi_{\mathcal{R}}) \nabla^i \overline{\psi} +  \phi \frac{1}{r^4} \right) dV\leq 0.
\end{align}

So let $\phi \in C_c^{\infty}(M_{\mathcal{R}})$ be a non-negative test function. By \eqref{SuperSolutionEstimate2} since $L_{\mathcal{R}}(\widetilde{\psi}) \leq -r^{-4}=L_{\mathcal{R}}(\psi_{\mathcal{R}})$
we have

\begin{align} \begin{split}
0 &\geq \int_{M_{\mathcal{R}}\cap M_{r_0}} \phi (L_{\mathcal{R}}(\widetilde{\psi})+ r^{-4}) dV = \int_{M_{\mathcal{R}}\cap M_{r_0}} \phi (L_{\mathcal{R}}(\overline{\psi})+ r^{-4}) dV \\ &=\int_{M_{\mathcal{R}}\cap M_{r_0}} \phi \left( \Delta_g \overline{\psi} + \frac{C}{r^2}  b_i(\nabla \psi_\mathcal{R}) \nabla^i \overline{\psi}  + \frac{1}{r^4}\right) dV \\
&=\int_{M_{\mathcal{R}}\cap M_{r_0}}  \left( -\langle \nabla \phi, \overline{\psi} \rangle_g + \phi \frac{C}{r^2}  b_i(\nabla \psi_\mathcal{R}) \nabla^i \overline{\psi}  + \phi \frac{1}{r^4}\right) dV + \int_{S_{r_0}} \phi \partial_r \overline{\psi} dA \\
&=\int_{M_{\mathcal{R}}\cap M_{r_0}}  \left( -\langle \nabla \phi, \overline{\psi} \rangle_g + \phi \frac{C}{r^2}  b_i(\nabla \psi_\mathcal{R}) \nabla^i \overline{\psi}  + \phi \frac{1}{r^4}\right) dV + \int_{S_{r_0}} \phi \partial_r \widetilde{\psi} dA
\end{split}
\end{align}
where we integrated by parts. Notice there is no boundary integral term on $\partial M$ since $\phi \in C_c^{\infty}(M_{\mathcal{R}})$ and the last term has a positive sign because the outward unit normal to $M_{\mathcal{R}}\cap M_{r_0}$ at $S_{r_0}$ points to infinity. 

Similarly using \eqref{SuperSolutionEstimate1} we have 
\begin{align} \begin{split}
0 &\geq \int_{M_{\mathcal{R}}\setminus M_{r_0}} \phi (L_{\mathcal{R}}(\overline{w})+ r^{-4}) dV = \int_{M_{\mathcal{R}}\setminus M_{r_0}} \phi (L_{\mathcal{R}}(\overline{\psi})+ r^{-4}) dV \\ &=\int_{M_{\mathcal{R}}\setminus M_{r_0}} \phi \left( \Delta_g \overline{\psi} + \frac{C}{r^2}  b_i(\nabla \psi_\mathcal{R}) \nabla^i \overline{\psi}  + \frac{1}{r^4}\right) dV \\
&=\int_{M_{\mathcal{R}}\setminus M_{r_0}}  \left( -\langle \nabla \phi, \overline{\psi} \rangle_g + \phi \frac{C}{r^2}  b_i(\nabla \psi_\mathcal{R}) \nabla^i \overline{\psi}  + \phi \frac{1}{r^4}\right) dV - \int_{S_{r_0}} \phi \partial_r \overline{\psi} dA \\
&=\int_{M_{\mathcal{R}}\setminus M_{r_0}}  \left( -\langle \nabla \phi, \overline{\psi} \rangle_g + \phi \frac{C}{r^2}  b_i(\nabla \psi_\mathcal{R}) \nabla^i \overline{\psi}  + \phi \frac{1}{r^4}\right) dV - \int_{S_{r_0}} \phi \partial_r \overline{w} dA
\end{split}
\end{align}
where again there is no surface integral on $S_{M_\mathcal{R}}$ because $\phi \in C_c^{\infty}(\mathcal{R})$ and the last term has a minus sign because the outward unit normal to $M_{\mathcal{R}}\setminus M_{r_0}$ on $S_{r_0}$ points away from infinity.  

Adding these two inequalities together then gives 
\begin{align}
    0\geq  \int_{M_{\mathcal{R}}} \left(-\langle \nabla\phi, \nabla \left( \overline{\psi} \right) \rangle_g + \phi \frac{C}{r^2}b_i(\nabla \psi_{\mathcal{R}})\nabla^i(\overline{\psi}) + \phi \frac{1}{r^4}\right)    dV + \int_{S_{r_0}} \phi \partial_r (\widetilde{\psi}-\overline{w}) dA
\end{align}
or
\begin{align}
    \int_{S_{r_0}} \phi \partial_r (\overline{w}- \widetilde{\psi}) dA\geq  \int_{M_{\mathcal{R}}} \left(-\langle \nabla\phi, \nabla \left( \overline{\psi} \right) \rangle_g + \phi \frac{C}{r^2}b_i(\nabla \psi_{\mathcal{R}})\nabla^i(\overline{\psi}) + \phi \frac{1}{r^4}\right)    dV.
\end{align}
But, by \eqref{NormalDerivativeEstimate} the left hand side is non-positive so therefore
\begin{align}
    0\geq  \int_{M_{\mathcal{R}}} \left(-\langle \nabla\phi, \nabla \left( \overline{\psi} \right) \rangle_g + \phi \frac{C}{r^2}b_i(\nabla \psi_{\mathcal{R}})\nabla^i(\overline{\psi}) + \phi \frac{1}{r^4}\right)    dV
\end{align}
and so indeed $\overline{\psi}$ is a weak supersolution to \eqref{MixedProblemSubSuper} for any $\mathcal{R}$. 

Since $\widehat{\psi}_{r_0}$ in the definition of $\overline{\psi}$ is a fixed function, we can take a sufficiently large $\lambda$ such that $\overline{\psi}>0$ everywhere. We'd like to show $\overline{\psi}\geq \psi_{\mathcal{R}}$ for all $\mathcal{R}$ to obtain a uniform $C^0$ bound for all the solutions.

Let $V_{\mathcal{R}}=\overline{\psi}-\psi_{R}$ on $M_{\mathcal{R}}$ so that $L(V_{\mathcal{R}})\leq 0$ on $M_{\mathcal{R}}$. In fact we have
\begin{align} \label{Contradiction}
    L_{\mathcal{R}}(V_{\mathcal{R}}) = L_{\mathcal{R}}(\overline{\psi}) - L_{\mathcal{R}}(\psi_{\mathcal{R}}) \leq -\frac{2}{r^4} + \frac{1}{r^4} = -\frac{1}{r^4} < 0 \quad \text{on} \quad M_{\mathcal{R}} \cap M_{r_0}.
\end{align}
Since $V_{\mathcal{R}}\in H^1(M_{\mathcal{R}}) \cap C^0(M_{\mathcal{R}})$ the weak maximum principle (see Theorem 8.1 in \cite{GilbargTrudinger}) gives
\begin{align}
    \inf_{M_{\mathcal{R}}} V_{\mathcal{R}} = \inf_{\partial M_{\mathcal{R}}} V_{\mathcal{R}}.
\end{align}
If $\inf_{\partial M_{\mathcal{R}}} V_{\mathcal{R}} \geq 0$ we're done. So suppose $V_{\mathcal{R}}<0$ somewhere, so that the infimum occurs necessarily somewhere on $\partial M_{\mathcal{R}}$.    Now, $\partial M_{\mathcal{R}}=\partial M \cup S_{\mathcal{R}}$. On $S_\mathcal{R}$ we have $V_{\mathcal{R}}>0$ because $\overline{\psi}>0$ and $\psi_{\mathcal{R}}=0$ on $S_{\mathcal{R}}$. Therefore, let us suppose that the infimum is reached at some point $x_0$ on $\partial M$. 

Now take $\overline{r}=\min\lbrace r_0, \mathcal{R} \rbrace$. Consider the set $\Lambda=M_{\overline{r}}$ and its interior $\Lambda^o$. We have $V_{\mathcal{R}}\in C^{2, \alpha}(\Lambda)$ and the infimum on this set is still reached at the same $x_0 \in \partial M$. We can now apply Lemma 3.4 from \cite{GilbargTrudinger} which gives that if $V_{\mathcal{R}}(x_0)<V_{\mathcal{R}}(x)$ for all $x\in \Lambda^o$ then necessarily
\begin{align}
    \frac{\partial V_{\mathcal{R}}}{\partial \nu}(x_0) < 0. 
\end{align}
But this is a contradiction, since $x_0 \in \partial M$ and our functions satisfy the zero Neumann condition there. (We remark that this boundary point lemma is proven for the Euclidean normal derivative in the given choice of coordinates, but it can be easily generalized to apply to the normal derivative with respect to the metric. Another way to see this is to take geodesic normal coordinates at $x_0$ so that the two normal derivatives match at that point in these coordinates.) 

Therefore, there must be some point $x\in \Lambda^o$ such that $V_{\mathcal{R}}(x_0)=V_{\mathcal{R}}(x)$. But then we can apply the strong minimum principle to conclude $V_{\mathcal{R}}$ is constant on $\Lambda \subset M_{\mathcal{R}}\cap M_{r_0}$. But this then contradicts \eqref{Contradiction}. 

Thus, we concldue that the infimum of $V_{\mathcal{R}}$ is not reached on $\partial M$. Therefore, it is reached on $S_\mathcal{R}$ where $V_{\mathcal{R}}$ is positive. Therefore
\begin{align}
    \inf_{M_{\mathcal{R}}} (\overline{\psi}-\psi_{\mathcal{R}})=\inf_{M_{\mathcal{R}}} V_{\mathcal{R}} = \inf_{\partial M_{\mathcal{R}}} V_{\mathcal{R}} \geq 0
\end{align}
so that
$\psi_{\mathcal{R}}\leq \overline{\psi}$. It is also elementary to show $0\leq \psi_{\mathcal{R}}$ giving us the uniform $C^0$ estimate
\begin{align}
    \label{C0Estimate} 0 \leq \psi_{\mathcal{R}}\leq \overline{\psi}.
\end{align}

\subsection{Global existence}

We can now use the uniform estimate \eqref{C0Estimate} to construct a convergent subsequence to obtain a global solution of \eqref{MainNeumannProblem}. Take a countable cover of $M$ consisting of smooth precompact open sets $\lbrace \Omega_n \rbrace$ (for example a countable collection of balls of Euclidean radius $1$ in our asymptotically flat coordinates intersected with $M$ will do). 

We can think of the functions $\psi_{\mathcal{R}}$ as solutions of the family of elliptic operators with uniformly bounded coefficients as in $\eqref{MixedProblemSubSuper}$ but now on each $\Omega_n$. For each $\Omega_n$ we have $\Omega_n \subset M_{\mathcal{R}}$ for all sufficiently large $\mathcal{R}$ due to the precompactness of the elements of the cover.

Therefore we can apply the usual $L^p$ estimates once again to obtain 
\begin{align}
    \Vert \psi_{\mathcal{R}} \Vert_{W^{2, p}(\Omega_n)} \leq C_n \left( \Vert r^{-4} \Vert_{{L^{p}(\Omega_n )}}  + \Vert \psi_{\mathcal{R}}\Vert_{L^{p}(\Omega_n 
 )}  \right)
\end{align}
where the elliptic constant $C_n$ depends on $\Omega_n$ but not on $\mathcal{R}$ due to the uniform boundedness of the coefficients in \eqref{MixedProblemSubSuper}. Combined with \eqref{C0Estimate} this then gives
\begin{align}
    \Vert \psi_{\mathcal{R}} \Vert_{W^{2, p}(\Omega_n)} \leq C_n \left( \Vert r^{-4} \Vert_{{L^{p}(\Omega_n )}}  + \Vert \overline{\psi}\Vert_{L^{p}(\Omega_n 
 )}  \right) \leq C'_n
\end{align}
for some constant $C'_n$ independent of $\mathcal{R}$.

Choosing suffieiently large $p$ and applying the Sobolev embedding theorem then gives a uniform estimate
\begin{align}
    \Vert \psi_{\mathcal{R}} \Vert_{C^{1, \gamma}(\overline{\Omega}_n)} \leq C_n''
\end{align}
for any $0<\gamma<1$ for some constant $C_n''$ independent of $\mathcal{R}$.
Using this estimate and looking at the $\psi_{\mathcal{R}}$ as solutions of 
\begin{align} 
        \begin{split}
            &\Delta_g \psi_{\mathcal{R}} = f_{\mathcal{R}}\quad \text{on} \quad \Omega_n \\
    f_{\mathcal{R}}&=- \frac{C}{r^2} \vert \nabla \psi_{\mathcal{R}} \vert_g  -\frac{1}{r^4} \in C^{0, \gamma}(\overline{\Omega}_n) \end{split} 
\end{align}
we have that the $f_{\mathcal{R}}$ are uniformly bounded in $C^{0, \gamma}(\overline{\Omega}_n)$. Combining this with the uniform estimate $\eqref{C0Estimate}$ and the usual Schauder estimates gives the uniform estimate
\begin{align}
    \Vert \psi_{\mathcal{R}} \Vert_{C^{2, \gamma}(\overline{\Omega}_n)} \leq C_n'''
\end{align}

Using the fact that the embedding $C^{2, \gamma} \hookrightarrow C^{2, \alpha}$ is compact for $0<\alpha < \gamma <1$, we can take a sequence $\psi_{\mathcal{R}_i}$ with $\mathcal{R}_i \rightarrow \infty$ which on any $\Omega_n$ has a convergent subsequence converging in $C^{2, \alpha}(\overline{\Omega}_n)$. Using the usual diagonal argument, we can then extract a subsequence converging on all $\Omega_n$ and thus obtain a subsequence converging on compact subsets of $M$ to a $C^{2, \alpha}(M)$ solution $\psi$ of \eqref{MainNeumannProblem} satisfying 
\begin{align}
    0\leq \psi \leq \lambda r^{-\beta} \quad \text{for} \quad r\geq r_0.
\end{align}

\subsection{Estimates at infinity}

Next, we want to show that the resulting solution $\psi$ is in fact in the weighted H\"{o}lder space $C^{2, \alpha}_{-1}(M)$. In \cite{HirschKazarasKhuri}, the authors refer to the methods of Proposition 3 in \cite{SchoenYau2} in order to obtain their desired fall-off conditions, but a careful look at that proposition only shows the solution to be in $C^{2, \alpha}_{-\beta}$ for $\beta \in (0, 1)$ which is not quite good enough for us. Thus, we make use of weighted elliptic estimates from \cite{Chaljub1979} and \cite{Weinstein_2005}.

Consider the region
\begin{align}
    \Omega=\lbrace x \in M: r(x)\geq \overline{r}_{0}>r_0 \rbrace. 
\end{align}
Take some point $x\in \Omega$ with $r(x)=r_x$ such that the ball of Euclidean radius $1$ centered at $x$ is contained in $\Omega$ so $B=B_1(x)\subset \Omega$. Notice, if we take $\overline{r}_0$ sufficiently large, then we have
\begin{align}
    \frac{1}{r_x-1} \leq \frac{2}{r_x}
\end{align}
so that on $B$ we have
\begin{align} \label{EstimateForRadiusInBall}
    \frac{1}{r(y)}\leq \frac{2}{r_x}
\end{align}
for all $y\in B$.

Now on each such ball, we can look at $\psi$ as a solution to the problem
\begin{align}  \label{DirichletProblem}
        \begin{split}
            L(v) \coloneqq &\Delta_g v + \frac{C}{r^2}  b_i(\nabla \psi_\mathcal{R}) \nabla^i v= -\frac{1}{r^4} \quad \text{on} \quad B 
        \end{split}
    \end{align}
which as before allows us to use the usual $L^p$ estimates giving
\begin{align}
    \Vert \psi \Vert_{W^{2, p}(B)} \leq C_e \left( \Vert r^{-4} \Vert_{{L^{p}(B)
  }}  + \Vert \psi \Vert_{L^{p}(B)  }  \right)
\end{align}
where the elliptic constant $C_e$ can be chosen so that it does not depend on the center of the ball due to the uniform boundedness of the coefficients (which is also a consequence of the asymptotic flatness). Since the ball has radius $1$ and is in $\Omega$ then we can estimate
\begin{align*}
    &\left(\int_B |r^{-4}|^p dV\right)^{1/p} \leq \left(\int_B |(2/r_x)^4|^p dV\right)^{1/p}=16\left( \frac{4}{3} \pi \right)^{1/p} r_x^{-4} \\
     &\left(\int_B |\psi|^p dV\right)^{1/p} \leq \left(\int_B |\lambda r^{-\beta}|^p dV\right)^{1/p} \leq \left(\int_B |\lambda 2^{\beta}r_x^{-\beta}|^p dV\right)^{1/p} =\lambda 2^\beta \left(  \frac{4}{3}\pi \right)^{1/p}r_x^{-\beta}.
\end{align*}
Putting these estimates together we then have
\begin{align}
    \Vert \psi \Vert_{W^{2, p}(B)} \leq C_e' r_x^{-\beta}
\end{align}
for some constant $C_e'$ depending on $\lambda, \beta, p$ but not the center of the ball $x$. For sufficiently large $p$ we can now apply the Sobolev embedding theorem, where the Sobolev constant does not depend on $x$ since we are looking at Euclidean balls of radius $1$. Therefore we get 
\begin{align} \label{AsymptoticEstimate1}
    \Vert \psi \Vert_{C^{1, \gamma}(\overline{B})} \leq C_e'' r_x^{-\beta}
\end{align}
where $\gamma=1-3/p$ and $C_e''$ doesn't depend on the center of the ball. Notice, this means $\partial_i \psi = O(r^{-\beta})$. 

Now, let us look at 
\begin{align}
f=- \frac{C}{r^2} \vert \nabla \psi \vert_g  -\frac{1}{r^4} \in C^{0, \gamma}(\overline{B})
\end{align}
which satisfies
\begin{align} \label{AsymptoticEstimate2}
    \Vert f \Vert_{C^{0, \gamma}(\overline{B})} \leq C_e''' r_x^{-\beta-2}
\end{align}
for some constant $C_e'''$ independent of $x\in \Omega$ by \eqref{AsymptoticEstimate1}. 

Now, take some $0<\alpha'<\gamma$ such that $\alpha'+\beta'<\beta$. Now, take any $y\in B_{1/4}(x)$. Let $\widetilde{r}(x, y)=\min\lbrace r(x), r(y)\rbrace < r_x$. We have
\begin{align} \begin{split}
    \sup_{y\in B_{1/4}(x)} (\widetilde{r}^{2+\beta'+\alpha'}(x, y))\frac{|f(x)-f(y)|}{|x-y|^{\alpha'}} &\leq r_x^{2+\beta'+\alpha'} \left( \sup_{y\in B_{1/4}(x)} \frac{|f(x)-f(y)|}{|x-y|^{\alpha'}} \right) \\
    & \leq r_x^{2+\beta'+\alpha'} \left( \sup_{y\in B_{1/4}(x)} \frac{|f(x)-f(y)|}{|x-y|^{\gamma}} \right) \\
    &\leq r_x^{2+\beta'+\alpha'} \left( C_e''' r_x^{-\beta-2}   \right) < \infty
    \end{split} 
\end{align}
Therefore by definition of weighted H\"{o}lder spaces (again, we refer the reader to \cites{EichmairHuangLeeSchoen2016, Chaljub1979, Weinstein_2005} for the definitions) we have $  f \in C^{0, \alpha'}_{-\beta'-2}(\overline{\Omega})$ and so in fact
\begin{align}
    f \in C^{0, \alpha'}_{-\beta'-2}(M).
\end{align}
Looking at $\psi$ as a solution of 
\begin{align} \label{WeightedNeumannProblem}\begin{split}
    &\Delta_g v = f \quad \text{on} \quad M \\
    &\left.\frac{\partial v}{\partial \nu}\right\vert_{\partial M} = 0 \end{split}
\end{align}
we can now use the results of \cite{Chaljub1979} to conclude that there exists a unique solution to \eqref{WeightedNeumannProblem} in $\widetilde{C}^{2, \alpha'}_{-\beta'}(M)$, which therefore is the same $\psi$ we found earlier. 
\begin{remark}
    Technically, the results of \cite{Chaljub1979} are proven for the case of complete asymptotically flat manifolds without boundary, but as mentioned in the footnote on page 11 of that reference, the results can be generalized to the case of manifolds with boundary. In particular, this is easy to see in our case since we can apply the Fredholm alternative to \eqref{WeightedNeumannProblem}.  
\end{remark}

Since we now have $\partial_i \psi = O(r^{-\beta'-1})$ this improves the asymptotics of $f$ giving 
\begin{align}
    f \in C^{0, \alpha'}_{-2-\beta}
\end{align}
for any $\beta \in (0, 1)$, yielding that $\psi \in \widetilde{C}^{2, \alpha'}_{-\beta}(M)$. Repeating this argument one final time again improves the asymptotics on $f$ and thus on $\psi$ giving 
\begin{align}
    \psi \in \widetilde{C}^{2, \alpha}_{-\beta}(M)
\end{align}
for any $\alpha, \beta \in (0, 1)$.

Now, notice if we take $\beta=-2/3$ then 
\begin{align}
f=- \frac{C}{r^2} \vert \nabla \psi \vert_g  -\frac{1}{r^4}  \in C^{0, \alpha}_{-3} \cap L^1(M).
\end{align}
We can now use the elliptic estimate from Lemma 2 of \cite{Weinstein_2005} which gives
\begin{align}
    \Vert \psi \Vert_{C^{2, \alpha}_{-1}(M)} \leq C \left(  \Vert f \Vert_{C^{0, \alpha}_{-3} \cap L^1(M)} + \Vert \psi \Vert_{C^{2, \alpha}_{-2/3}(M)}   \right)
\end{align}
for some constant $C$. Therefore indeed $\psi \in C^{2, \alpha}_{-1}(M)$ as desired. 

\section{Possible Further Applications} \label{Conclusion}

One possible application of the methods of this paper is to the deformation of charged initial data sets. One can consider an initial data set $(M, g, k, E, B)$ where $E$ and $B$ are the electric and magnetic fields. Then the charged dominant energy condition takes the form
\begin{align}
    \mu_{EM} \geq |J_{EM}|_g
\end{align}
where $\mu_{EM}$ and $J_{EM}$ denote the energy and momentum densities respectively with the contributions from the electromagnetic fields subtracted off. As far as we have seen, no general result about deforming such initial data sets to the strict charged dominant energy condition have been published, and the results of the current paper should readily be able to handle this case.

Some other possible applications are to the Penrose inequality, which is one of the most important open problem in mathematical general relativity \cite{Penrose}, which relates the total mass of a spacetime to the area of the black holes present in the spacetime. The mathematically precise formulation of the conjecture for asymptotically flat initial data sets is believed to be
\begin{align} \label{PenroseInequality}
    E_{ADM} = \sqrt{  \frac{A}{16\pi}} 
\end{align}
where $A$ is the total area of the outermost area enclosure of future/past apparent horizons present in the initial data sets. The inequality has been proven in the so-called Riemannian case where $k=0$ and the dominant energy condition reduces to non-negative scalar curvature, and the outermost apparent horizon reduces to an outermost minimal surface \cites{Bray, HuiskenIlmanen}. Various generalizations to include charge and angular momentum have been proposed and proven under certain conditions on $k$ \cites{KhuriWeinsteinYamada, KhuriWeinsteinYamada1, JaraczKhuri, JaraczCylindricalPenrose, KhuriSokolowskyWeinstein}, but even the simplest version \eqref{PenroseInequality} for general $k$ remains open. While some general approaches have been proposed \cite{BrayKhuriPDE}, it has turned out some of these approaches do not work, as for example the proposed Jang/zero-divergence system \cite{Jaracz_2023_ZeroDivergence}.

If the Penrose conjecture turns out to be true, then the dominant energy condition is a key ingredient. In fact the heuristic physics argument which leads to the Penrose inequality relies on the black hole area theorem which holds under the weaker assumption of either the weak energy condition or the null energy condition. The dominant energy condition implies the weak energy condition which in turn implies the null energy condition, but not conversely. However, there exist counterexamples to both the positive mass theorem and the Penrose inequality under the assumption of the weak energy condition \cite{JaraczWEC}. It remains unclear what causes this breakdown between the physical and mathematical arguments. 

The most powerful approach to the Penrose inequality was the conformal approach of Bray \cite{Bray} where the flow was defined by 
\begin{align} \label{ConformalFlowBray}
    \begin{split}
       u_t &= 1 + \int_0^t v_s ds \\
        \Delta_{g} v_t &= 0 \quad \text{on} \quad M_t, \\
        v_t \equiv 0 \quad \text{on} \quad M\setminus M_t, \quad v_t \vert_{\partial M_t}&=0, \quad \lim_{|x|\rightarrow \infty} v_t(x)=-e^{-t}.
    \end{split}
\end{align}
Here, $M_t$ is the region outside the outermost minimal surface in $(M, g_t)$ where $g_t=u_t^4 g$. This conformal flow preserved the non-negativity of scalar curvature. 

Similarly, looking at \eqref{ImportantEquation} we see that if we have a family of conformal factors which satisfy 
\begin{align} \label{ImportantEquation2}
    \Delta_g u_t =  -|k_{ij}\nabla^j u_t|_g
\end{align}
then this family of metrics will preserve the dominant energy condition. In fact, we could have studied this type of equation instead, but using the estimate \eqref{Estimatek} proved to be more convenient in the present work. In addition, if we assume
\begin{align}
     u_t &= 1 + \int_0^t v_s ds
\end{align}
so that $u_t'=v_t$ we see that \eqref{ImportantEquation2} is true for $t=0$. Then letting $k_{ij}\nabla^j u_t= Y_i$ and differentiating \eqref{ImportantEquation2} with respect to $t$ gives
\begin{align} \label{ImportantEquation3}
    \Delta_g v_t = -\frac{Y^i k_{ij}}{|Y_i|} \nabla^j v_t.
\end{align}
In the special case where $k\equiv 0$ this reduces to \eqref{ConformalFlowBray}. It is unclear if such a flow could be used to tackle the general case of the Penrose inequality, but it looks like a promising possibility.

\appendix

\bibliography{ref.bib}

 \footnotesize

  J.S.~Jaracz, \textsc{Department of Mathematics, Texas State University,
    San Marcos, TX 78666}\par\nopagebreak
  \textit{E-mail address} \texttt{jaracz@txstate.edu}

\end{document}